\documentclass[twocolumn,superscriptaddress,longbibliography]{revtex4-1}
\pdfoutput=1
\usepackage{amsfonts,amsmath,amssymb}
\usepackage{enumerate}
\usepackage{hyperref}
\newcommand{\be}{\begin{equation}}
\newcommand{\ee}{\end{equation}}

\newcommand{\bea}{\begin{eqnarray}}
\newcommand{\eea}{\end{eqnarray}}

\newcommand{\bal}{\begin{align}}
\newcommand{\eal}{\end{align}}

\newcommand{\bes}{\begin{subequations}}
\newcommand{\ees}{\end{subequations}}

\newcommand{\nn}{\nonumber}

\newcommand{\ra}{\rightarrow}

\begin{document}

\title{Surface/state correspondence and bulk local operators \\
in pp-wave holography}
\author{Nakwoo Kim}
\email{nkim@khu.ac.kr}
\affiliation{Department of Physics
and Research Institute of Basic Science,  Kyung Hee University, 
 Seoul 02447,
  Korea}
  \affiliation{School of Physics, Korea Institute for Advanced Study, Seoul 02455, Korea}
\author{Hyeonjoon Shin}
\email{nonchiral@gmail.com}
\affiliation{School of Physics, Korea Institute for Advanced Study, Seoul 02455, Korea}

\begin{abstract}
We apply the surface/state correspondence proposal of Miyaji {\it et al.} to IIB pp-waves and propose that the bulk local operators should be instantonic D-branes. In line with ordinary AdS/CFT correspondence, the bulk local operators in pp-waves also create a hole, or a boundary, in the dual gauge theory as pointed out by H. Verlinde, and by Y. Nakayama and H. Ooguri. We also present simple calculations which illustrate how to extract the spacetime metric of pp-waves from instantonic D-branes in boundary state formalism.
\end{abstract}

\maketitle

\section{Introduction}
Recently there appeared an interesting proposal called surface/state correspondence
\cite{Miyaji:2014mca,Miyaji:2015yva,Miyaji:2015fia} which aims to generalize the idea and applicability of holography \cite{'tHooft:1993gx,*Susskind:1994vu, *Maldacena:1997re}. It is partly motivated by the observation that a particular tensor network prescription, called MERA (Multiscale Entanglement Renormalization Ansatz) \cite{Vidal:2007hda}, apparently realizes the emergence of holographic spacetime through a real space renormalization \cite{Swingle:2009bg}. 
The idea of surface/state correspondence is based on the holographic entanglement entropy conjecture \cite{Ryu:2006bv,*Ryu:2006ef}, and extends it by assigning a density matrix not only to a spacelike minimal surface ending on the boundary but also to any open or closed surface in the bulk. Moving and deforming the surface in the bulk should correspond to performing the coarse-graining procedure: dubbed as ``disentangler'' and ``isometry'' in MERA literature. If possible, this generalized holography relation  can be also applied to bulk geometries without a boundary, including the Minkowski and de Sitter spacetime. In particular, when the surface shrinks to a point the dual state should be bulk local operators which have been studied in {\it e.g.} Refs.\cite{Hamilton:2006fh,*Kabat:2011rz,*Kabat:2014kfa,*Kabat:2015swa}.

In this note we consider the maximally supersymmetric IIB pp-wave \cite{Blau:2001ne}, and its gauge/gravity correspondence by Berenstein, Maldacena and Nastase (BMN) \cite{Berenstein:2002jq}. The advantage is obviously twofold. The string spectrum is exactly solvable in the lightcone gauge, and the proposed mapping to large R-charge operators has been extensively studied and confirmed. On the other hand, 
just like the flat background the pp-wave does not have a clear boundary and the
notion of holography has remained rather ambiguous \cite{Das:2002cw,*Kiritsis:2002kz,Leigh:2002pt}. Thus we reckon the pp-wave/BMN 
relation provides an excellent setup to test the surface/state correspondence in more detail and also in the full string theory, potentially including even the string interaction through string field theory \cite{Berenstein:2002sa}.

Our main point is that instantonic D-branes are the ``surfaces'' which are mapped to gauge theory ``states'' through the BMN relation to gauge operators. It is probably important to emphasize here that the brane configurations we will consider
are {\it not} defined in Wick-rotated euclidean bulk spacetime. The D-brane worldvolume is euclidean, and they 
occupy a spacelike subspace in pp-wave background. Back to the surface/state correspondence, 
in particular for the bulk local operators we use D-instantons. D-instantons probing the bulk geometry is a very well known idea in AdS/CFT: they are dual to Yang-Mills instantons, of which the moduli space contains $AdS_5$ precisely \cite{Bianchi:1998nk}. As usual, it is expected that the supersymmetry and modular invariance of 1/2-BPS D-instantons will protect them against perturbative corrections and string interactions.

We take a modest first step in this note. Following mainly the discussions in Refs.\cite{Miyaji:2015fia,Verlinde:2015qfa,Nakayama:2015mva}, we first discuss how the pp-wave symmetry algebra should constrain the candidate bulk local operators, and also illustrate how to extract the information of the pp-wave metric from D-instantons. Along the way we also argue that the D-instantons as bulk local operators are ``boundary creating'' operators on the boundary, just as proposed in Refs.\cite{Verlinde:2015qfa,Nakayama:2015mva}. We conclude with discussions on possible future avenues.

\section{IIB pp-wave and BMN correspondence}
In this section we give a brief review of IIB pp-wave geometry \cite{Blau:2001ne} and its
gauge/gravity duality relation \cite{Berenstein:2002jq}. 
The metric of IIB maximally supersymmetric pp-wave is given as ($I=1,2,\cdots,8$)
\be
ds^2 = 2dx^+ dx^- - \mu^2 (x^I)^2(dx^+)^2 + dx^I dx^I \, . 
\ee
Note that the apparent $SO(8)$ symmetry of the metric is broken 
into $SO(4)\times SO(4)$ due to the nonvanishing RR 5-form 
$F_{+1234}=F_{+5678}=2\mu$. This background is related to $AdS_5\times S^5$ through the Penrose limit, where one takes a null geodesic, and blows up the spacetime around it. Morally speaking the first $\mathbb{R}^4$ is understood as Euclidean spacetime, and the second $\mathbb{R}^4$ is the (infinitely magnified) four-dimensional space around a great circle in $S^5$. One comment is that to obtain this nontrivial pp-wave the null geodesic should be extended 
in both $AdS_5$ and $S^5$. If the null curve is totally inside $AdS_5$, the resulting background is just the trivial
Minkowski background without any interesting gauge/gravity correspondence.

The total bosonic symmetry group is 30 dimensional: there are kinematical generators $P^+,P^I,J^{+I},M^{ij},M^{i'j'}$
and a dynamical generator $P^-$ (here and below 
$i,j=1,2,3,4$ and $i',j'=5,6,7,8$). More explicitly $P^{\pm}=i\partial_{\mp}$, and
\bea
P^I &=& -i( \cos(\mu x^+)\partial_I + \mu \sin(\mu x^+) x^I \partial_-) \, , 
\\ 
J^{+I} &=& -i(\mu^{-1} \sin(\mu x^+)\partial_I - \cos(\mu x^+) x^I \partial_-) . 
\eea
$M^{ij},M^{i'j'}$ are the usual rotation generators of $SO(4)\times SO(4)$.
The relevant commutation
relations are 
\begin{align}
 [P^-, P^I] &= -i \mu^2 J^{+I}, 
\quad
[P^-, J^{+I}] = i P^I, 
\nn
\\
[P^I, J^{+J}] &=  -i\delta^{IJ} P^+ . 
\end{align}
$P^+$ is a central element and commutes with all other
generators. It is easy to see that $P^-,P^I,J^{+I}$ 
comprise Heisenberg algebras when we define $a^I=(\mu J^{+I}+iP^I)/\sqrt{2\mu P^+}$ and $a^{I\dagger}=(\mu J^{+I}-iP^I)/\sqrt{2\mu P^+}$. In that case $-P^-/\mu$ becomes the
number operator, since $[P^-,a^I]=\mu a^I$. The isometry generators can be of course more explicitly written in terms of string modes in lightcone gauge which will be explained below.

The proposal of Ref. \cite{Berenstein:2002jq} by Berenstein, Maldacena and Nastase (BMN) relates a certain class of long operators in ${\cal N}=4$ supersymmetric Yang-Mills (SYM) theory to the string spectrum in pp-wave background. The relevant operators, which we will call BMN operators, carry a large $U(1)_R$ charge $J$, so the conformal dimension $\Delta$ is also large. The BMN correspondence equates $-P^-/\mu=\Delta-J$ as a duality relation.

 The Green-Schwarz superstring theory of pp-waves in lightcone gauge turns out to be a free theory of 8 massive scalars and 8 massive fermions \cite{Metsaev:2002re}. 
The quantum spectrum starts with a unique vacuum state $|0;p^+\rangle$, where $p^+$ is lightcone momentum and a free parameter. There is a tower of string modes: first the zero modes $\alpha^I_0,\theta_0$ and their conjugates. For $n\ge 1$, $\alpha^I_n,\tilde\alpha^I_n$ are the right and the left moving bosonic modes in $SO(8)$ vector representation, while $\theta_n,\tilde\theta_n$ are the right and the left moving fermionic oscillators in spinor representation of $SO(8)$. The string vacuum 
 $|0;p^+\rangle$ is the Fock vacuum with lightcone energy $P^-=0$. Raising the level of a bosonic or a fermionic zero mode by one will increase
 $\Delta-J$ by one precisely. On the other hand, adding the occupation number of the $n$-th mode by one increases $\Delta-J$ by $\sqrt{1+n^2/(\alpha' p^+ \mu)^2}$. The usual level matching condition also applies, and the total occupation number of left- and right-moving modes should be the same for physical states.

The string spectrum beautifully matches that of the 
large R-charge operators in ${\cal N}=4$ SYM.  
    The vacuum state of the string theory $|0;p^+\rangle$ corresponds to
${\rm Tr}(Z^J)$, where $Z$ is one of the complex scalar fields in ${\cal N}=4$ SYM with $\Delta=J=1$. These are the only kind of ${\cal N}=4$ operators which satisfy $\Delta=J$ classically and also in the interacting theory. 
The eight bosonic fields on the worldsheet correspond to: 4 remaining scalar fields $\phi^i \,\,(i=1,\cdots,4)$ of SYM, and covariant derivatives $D_\mu$ along 4 Euclidean spacetime directions. And the eight fermionic directions come from the half of the set of fermion fields in ${\cal N}=4$ SYM with $\Delta=3/2,J=1/2$. The BMN excited states are long operators where the aforementioned ``impurity'' are inserted in the ``string'' of $Z$ fields. Excitation by zero modes are constructed as totally symmetric combinations: for instance (up to a normalization constant),
\be
a^{i\dagger}_0 a^{j'\dagger}_0 |0;p^+\rangle
\longleftrightarrow
\sum_l {\rm Tr} (Z^l (D_i Z) Z^{J-l} \phi^{j'})
\ee
and in an analogous way with fermions. These totally symmetrized combinations are also 1/2-BPS just as ${\rm Tr}Z^J$, so the conformal dimensions are protected. It agrees with the string side computation that $\Delta-J$ is exactly integral. For non-zero modes, the BMN proposal is that one should now consider impurities with ``momentum''. It goes like for instance
\begin{align}
& a^{i'\dagger}_n a^{j'\dagger}_n |0;p^+\rangle
\nn\\
&\longleftrightarrow
\sum_l e^{2\pi i nl/J} {\rm Tr}(Z^l \phi^{i'}  Z^{k} \phi^{j'} Z^{J-k})
\, . 
\end{align}
Then the string spectrum predicts that the SYM operator should receive quantum corrections for conformal dimension so that $\Delta-J=2\sqrt{1+n^2/(\alpha' p^+ \mu)^2}$. This is justified from the explicit perturbative computations in SYM, upon the duality identification \cite{Berenstein:2002jq}
\be
\frac{1}{(\alpha'p^+\mu)^2} = \frac{g^2_{YM}N}{J^2} \, . 
\ee
This proposal was extensively analyzed and confirmed to higher orders in perturbation theory, for a review see e.g. \cite{Sadri:2003pr}.

\section{Bulk local operators as D-instantons}
Let us now consider how the isometry algebra should restrict the putative BMN operator assignment in the bulk of IIB pp-wave. We will employ the closed string theory states here mainly because of notational simplicity: thanks to BMN proposal one can 
immediately write down the dual gauge theory operators. The maximally supersymmetric pp-wave is homogeneous, so we may start with any convenient point: let us choose $x^+=x^-=x^I=0$. Obviously the little group is then generated by $J^{+I},M^{ij},M^{i'j'}$ 
. The candidate string states are easily constructed. For the Fock vacuum $|0\rangle$ (in string theory realization of course this is the unique string ground state $|0;p^+\rangle$) satisfying $a^I|0\rangle=0$, the ``position'' eigenstate
$J^{+I}|B\rangle=0$ is given by 
\be
|B\rangle = e^{
\tfrac{1}{2}\sum_I (a^{I\dagger})^2 
} |0\rangle \, . 
\label{dinstanton}
\ee

In the string theory realization of course the symmetry requirement alone does not lead to a unique string state. There are fermionic modes as well, but more importantly the kinematical generators only involve the zero-modes and we may consider excitation by higher string modes. This is akin to the conformal field theory consideration in Refs \cite{Miyaji:2015fia,Verlinde:2015qfa,Nakayama:2015mva}, that in general the symmetry requirement for boundary states is satisfied by a linear combination of Ishibashi states. It is a natural guess that in string theory the bulk local state $|B\rangle$ should be D-instantons constructed as coherent states in closed string theory. Indeed, IIB pp-waves do allow 1/2-BPS D-instantons \cite{Gaberdiel:2002hh} which do include fermionic and other higher stringy mode excitations, and they are in general dual to long operators in ${\cal N}=4$ super Yang-Mills with fermion and gauge field insertions. We propose the
operators thus obtained to be the bulk local operators. Summarizing, the idea is that we have bulk local operator expressions in closed string theory constructed in  \cite{Gaberdiel:2002hh}, which then can be mapped SYM operators through BMN proposal. 

A couple of comments are in order here. We first note that the BMN dual of \eqref{dinstanton} are in general highly non-local in the dual field theory point of view, since for instance acting on $a^{i'\dagger}_n$ means insertion of a covariant derivative. And D-branes are non-perturbative objects (solitons) in string theory, so the bulk local operators should be understood as non-perturbative states in SYM. Secondly, in string theory there is a distinction between D-branes and anti D-branes. In the original proposal of surface/state correspondence \cite{Miyaji:2014mca} the bulk local operators are given as closed surface shrunk to a point. The lesson here is:
as one tries to embed it in string theory we need to consider not only the shape but also the orientation of the surface which can be related to the charge of the D-instantons.

For the standard AdS/CFT correspondence, it was argued that the local bulk operators are mapped to {\it boundary creation} operators through a unitary transformation \cite{Miyaji:2015fia,Verlinde:2015qfa,Nakayama:2015mva}. It was in particular emphasized in \cite{Nakayama:2015mva} that the mapping is mediated through dilatation (i.e. Hamiltonian in the radial quantization) by the imaginary unit. The spherical boundaries of the holes created on the CFT side are put at a constant time slice.
   
For the holography of pp-waves, we recall that the light-cone coordinate $x^+$ becomes time both on the string worldsheet and presumably also for the dual BMN sector. One way to see it is to rewrite the
metric as follows \cite{Leigh:2002pt}.
\bea
ds^2 &=& 2dx^+ dx^- + dr^2_1 + r^2_1( - \mu^2(dx^+)^2 + d\Omega^2_{3(1)}) 
\nn\\
&+& 
dr^2_2+ r^2_2( - \mu^2(dx^+)^2 + d\Omega^2_{3(2)})   
\, . 
\label{metric2}
\eea
It is implied here that the radial coordinates of $\mathbb{R}^4\times \mathbb{R}^4$ are the holographic coordinates, and the boundary includes two copies of $\mathbb{R}\times S^3$. We note that in the surface/state correspondence \cite{Miyaji:2014mca,Miyaji:2015fia} the holes are codimension-two spacelike hypersurfaces in the bulk. From the rearrangement
in \eqref{metric2} we infer that it is a natural guess that the hole
is at $x^+=0$, extended along $x^-$, and at $r=\infty$. This subspace is invariant under the same symmetry algebra of the point of our interest: $J^{+I}$ and $SO(4)\times SO(4)$ rotations. It is amusing to see that, in general, 
\be
|\mbox{A boundary at } x^+=x^+_0\rangle = e^{i x_0 P^-}|B\rangle \, . 
\ee
This is the central point of this short note:
namely, the bulk local operator and the spherical boundary are again mapped through the time evolution in the boundary theory in pp-wave holography.

Thanks to the relative simplicity of pp-wave geometry, it is easy to consider planar hypersurface operators in the bulk. If the surface is localized along $I\in {\cal D}$ and extended along $I\in {\cal N}$ directions, it is invariant under $P^I$ for $I\in {\cal N}$ and $J^{+I}$ for $I\in {\cal D}$. The reduction of the rotation group $M_{ij},M_{i'j'}$ should be done in the obvious way. Then the associated bulk operators in the bosonic zero sector should be
\be
|B\rangle = 
e^{
+\tfrac{1}{2}\sum_{I\in {\cal D}} (a^{I\dagger})^2 
-\tfrac{1}{2}\sum_{I\in {\cal N}} (a^{I\dagger})^2 
} |0\rangle \, . 
\ee
The ambiguity with respect to stringy modes and fermions should be fixed using worldsheet modular invariance and supersymmetry. As far as we know the analysis of modular invariance for pp-wave instantons is not complete, although it is worked out for a number of cases in Refs \cite{Bergman:2002hv,Gaberdiel:2002hh}. 
But it is known how unbroken supersymmetry restricts the possible D-brane configurations, depending on how many legs they have in each $\mathbb{R}^4$ \cite{Skenderis:2002vf}. According to the analysis in \cite{Gaberdiel:2002hh}, there exist supersymmetric instantonic D-branes of type $(0,0),(2,2),(4,0),(0,4),(4,4),(1,1),(3,3)$ where $(p,q)$ refers to the number of legs in each $\mathbb{R}^4$. In Ref.\cite{Gaberdiel:2002hh} 
the authors studied explicitly $(0,0),(4,0),(0,4)$ branes. Once they are constructed  in closed string theory the identification as operators in BMN sector is of course  straightforward.

Let us here consider specifically the case of $(0,4)$ (D3) and $(4,4)$ (D7) since they are easier to understand holographically. The $(0,4)$ case can be understood as a local operator in SYM spacetime, but in the bulk it covers a four-dimensional subspace in $S^5$ transverse to a great circle in terms of the original AdS/CFT correspondence. In other words, it is a local operator in $AdS_5$ but a hypersurface in $S^5$, before we take the Penrose limit. The preserved symmetries are $J^{+i},P^{i'}$ in addition to $M^{ij},M^{i'j'}$. On the holographic screen $r\rightarrow\infty$ at $x^+=0$, 
the same symmetry is preserved for a space which is a product of: $S^3$ in the first $\mathbb{R}^4$, the entire second $\mathbb{R}^4$, and the real line of $x^-$. In short, an operator which is local in the first $\mathbb{R}^4$ but non-local in the second $\mathbb{R}^4$ in the bulk is boundary creating operator only in the first $\mathbb{R}^4$ This is a natural result. The causal cone (in the sense of AdS/MERA \cite{Swingle:2009bg}) of a point in the bulk extended to UV is obviously a ball in boundary, but the causal cone of the entire bulk should cover the entire boundary spacetime. And there is no boundary to be created for $(4,4)$ D7-branes obviously.

Let us now go back to the bulk local state and the D-instanton, and check they have the properties which are naturally expected from the particle dynamics in pp-waves.
The first result we would like to reproduce is the solution of Klein-Gordon equation in pp-wave. The $(x^I)^2 (dx^+)^2$ term in the metric introduces a harmonic potential in $\mathbb{R}^8$, and one can easily obtain the eigenmodes of the $D=10$ massless scalar wave equation $\Box \Psi=0$ as follows (see e.g. \cite{Leigh:2002pt})
\begin{align}
\Psi_{n_I}
&= e^{ip^+ x^-} e^{i P^- ( n_I )x^+ } 
\prod_I \psi_{n_I}(x^I)
\label{eigenm}
\\
\psi_n (x) &=\frac{1}{\pi^{1/4}\sqrt{2^n n!}}e^{-\mu x^2/2} H_{n} (\sqrt{\mu} x)
\\
P^- (n_I) &= -\mu  \sum_I (n_I+1/2)
\end{align}
On the other hand, one can compute the overlap of local states at two different bulk points when we recall $|B\rangle$ is basically a position ($J^{+I}$) eigenstate.
\begin{align}
&\langle B | e^{i(p^+ x^- + P^- x^+)} e^{-ix^I P^I}| B \rangle
\nn
\\
&= e^{ip^+ x^-}
\sum_{{\rm even} \,\, n^I} e^{iP^-(n^I)x^+}
\psi_{n_I}(0) \psi_{n_I}(\sqrt{\mu}x) \, . 
\end{align}
Being a linear combination of the eigenmodes \eqref{eigenm} it is clear that the above expression satisfies the scalar equation in the bulk. 

The overlap function of a D-instanton vs. an anti D-instanton can be of course calculated in the full lightcone superstring theory, and the result can be found 
in Ref.\cite{Gaberdiel:2002hh}. Adjusting to our notation,
\begin{align}
{\cal A} (\Delta x^+; {\bf x}_1, {\bf x}_2) 
&=
h_0(\Delta x^+;{\bf x}_1,{\bf x}_2)
\nn\\
 &  \times (\text{non-zero mode contribution}) \, , 
\end{align}
where the zero-mode part contribution $h_0$ is given by 
\begin{align}
&h_0(\Delta x^+,  {\bf x}_1, {\bf x}_2)
\nn\\
&= \exp \left( - \frac{m (1+q^m) ({\bf x}_1^2 + {\bf x}^2_2)}{2 (1-q^m)}
               + \frac{2 m q^{m/2} {\bf x}_1 \cdot {\bf x}_2}{(1-q^m)}
        \right) \, . 
\end{align}
The parameters here are defined as follows,
\begin{align}
m =  p^+\mu \,, \quad
q= e^{- 2\Delta x^+/ p^+} \, . 
\end{align} 
In the above $\Delta x^+$ is the difference of $x^+$ coordinate, and
${\bf x}_1,{\bf x}_2$ are position vectors in the transverse $\mathbb{R}^8$.
We expect to be able to ``read off'' the bulk metric when we take the limit
$\Delta x^+\ra 0$ and $\Delta{\bf x}\equiv{\bf x}_2-{\bf x}_1\ra 0$.
In that limit, setting ${\bf x}\equiv({\bf x}_1+{\bf x}_2)/2$,
\begin{align}
1-h_0 
\simeq \frac{p^+}{2} \left[ 
          \left(\frac{\Delta {\bf x}^{}}{\Delta x^+}\right)^2 
          +\mu^2 {\bf x}^2  \right]
         \Delta x^+ \,.
\end{align}
We note that this expression is consistent with the pp-wave metric, recalling the dispersion relation $-2p^+ P^- = {\bf p}^2 +\mu^2 {\bf x}^2$ and considering the lightcone time evolution operator in Euclidean signature $e^{iP^-\Delta x^+}$.

\section{Discussion}
We have studied the IIB pp-wave D-instantons in the light of the surface/state correspondence \cite{Miyaji:2015yva}.
Admittedly we only addressed the kinematics and zero-mode sector of the string theory so far to check the surface/state
 relation, and it is desirable to probe genuinely stringy aspects which are included in the construction of \cite{Gaberdiel:2002hh}. 
It will be also very nice if one can find MERA-type unitary transformations between the various instantonic D-brane constructions in \cite{Gaberdiel:2002hh}.

The pp-wave/gauge theory correspondence involves a subsector of gauge theory with large conformal dimensions
and R-charges \cite{Berenstein:2002jq}. We may ask if we can take a similar limit directly in the construction of
surface/state correspondence for general CFTs in \cite{Miyaji:2014mca,Miyaji:2015yva,Miyaji:2015fia}. A requirement
is that we should extend the class of CFTs with global symmetries and also append the $AdS_3$ space by a certain
extra internal space. This is because it is only when we take the large conformal dimension and large global charge
limit simultaneously that we obtain pp-waves with nontrivial duality relation with a subsector of gauge theory.
In principle one can take the limit with IIB strings in $AdS_5\times S^5$ too, but the usual difficulty of string quantization 
in Ramond-Ramond background applies. It will be very interesting if we can take a concrete example of 
$AdS_3\times S^3\times T^4$, work out the surface/state relation, and take the Penrose limit explicitly and
find the resulting bulk local operators are indeed instantonic D-branes.

Probably the biggest difference between the holographic entangling surface and D-branes is that the latter is dynamical. In other words they not only represent a density matrix, but can also induce a change in vacuum density matrix. It will be very interesting if we can relate the gravitational backreaction of D-instantons using the recent proposals on holographic Fisher information metric \cite{MIyaji:2015mia,*Lashkari:2015hha}.

The study of M-theory pp-waves and their relation to a massive matrix theory \cite{Berenstein:2002jq} is also an interesting subject. The kinematical consideration should straightforwardly carry through, in terms of the abelian part of the BMN matrix model. On the other hand, there is no systematic way of constructing M-branes in the BMN matrix level at quantum level. And according to the supergravity probe analysis in Ref.\cite{Kim:2002tj} there are no supersymmetric instantons in $D=11$ pp-waves, so we have a more challenging problem.
\begin{acknowledgments}
We thank Jung Hoon Han, Jeong-Hyuck Park, and Sang-Heon Yi for discussions.
This work was supported by  
NRF grants funded by the Korea government 
with grant No. 2010-0023121, 2012046278 (N.K.) and NRF-2012R1A1A2004203, 
2012-009117, NRF-2015R1A2A2A01004532 (H.S.). This work was completed while
NK was visiting CERN for ``CERN-CKC TH Institute on Duality Symmetries in
String and M-Theories''. NK is
grateful to CERN for hospitality, and to CERN-Korea theory collaboration committee 
for supporting the workshop. 
\end{acknowledgments}
\bibliography{papers}

\begin{thebibliography}{31}%
\makeatletter
\providecommand \@ifxundefined [1]{%
 \@ifx{#1\undefined}
}%
\providecommand \@ifnum [1]{%
 \ifnum #1\expandafter \@firstoftwo
 \else \expandafter \@secondoftwo
 \fi
}%
\providecommand \@ifx [1]{%
 \ifx #1\expandafter \@firstoftwo
 \else \expandafter \@secondoftwo
 \fi
}%
\providecommand \natexlab [1]{#1}%
\providecommand \enquote  [1]{``#1''}%
\providecommand \bibnamefont  [1]{#1}%
\providecommand \bibfnamefont [1]{#1}%
\providecommand \citenamefont [1]{#1}%
\providecommand \href@noop [0]{\@secondoftwo}%
\providecommand \href [0]{\begingroup \@sanitize@url \@href}%
\providecommand \@href[1]{\@@startlink{#1}\@@href}%
\providecommand \@@href[1]{\endgroup#1\@@endlink}%
\providecommand \@sanitize@url [0]{\catcode `\\12\catcode `\$12\catcode
  `\&12\catcode `\#12\catcode `\^12\catcode `\_12\catcode `\%12\relax}%
\providecommand \@@startlink[1]{}%
\providecommand \@@endlink[0]{}%
\providecommand \url  [0]{\begingroup\@sanitize@url \@url }%
\providecommand \@url [1]{\endgroup\@href {#1}{\urlprefix }}%
\providecommand \urlprefix  [0]{URL }%
\providecommand \Eprint [0]{\href }%
\providecommand \doibase [0]{http://dx.doi.org/}%
\providecommand \selectlanguage [0]{\@gobble}%
\providecommand \bibinfo  [0]{\@secondoftwo}%
\providecommand \bibfield  [0]{\@secondoftwo}%
\providecommand \translation [1]{[#1]}%
\providecommand \BibitemOpen [0]{}%
\providecommand \bibitemStop [0]{}%
\providecommand \bibitemNoStop [0]{.\EOS\space}%
\providecommand \EOS [0]{\spacefactor3000\relax}%
\providecommand \BibitemShut  [1]{\csname bibitem#1\endcsname}%
\let\auto@bib@innerbib\@empty
\bibitem [{\citenamefont {Miyaji}\ \emph
  {et~al.}(2015{\natexlab{a}})\citenamefont {Miyaji}, \citenamefont {Ryu},
  \citenamefont {Takayanagi},\ and\ \citenamefont {Wen}}]{Miyaji:2014mca}%
  \BibitemOpen
  \bibfield  {author} {\bibinfo {author} {\bibfnamefont {Masamichi}\
  \bibnamefont {Miyaji}}, \bibinfo {author} {\bibfnamefont {Shinsei}\
  \bibnamefont {Ryu}}, \bibinfo {author} {\bibfnamefont {Tadashi}\ \bibnamefont
  {Takayanagi}}, \ and\ \bibinfo {author} {\bibfnamefont {Xueda}\ \bibnamefont
  {Wen}},\ }\bibfield  {title} {\enquote {\bibinfo {title} {{Boundary States as
  Holographic Duals of Trivial Spacetimes}},}\ }\href {\doibase
  10.1007/JHEP05(2015)152} {\bibfield  {journal} {\bibinfo  {journal} {JHEP}\
  }\textbf {\bibinfo {volume} {1505}},\ \bibinfo {pages} {152} (\bibinfo {year}
  {2015}{\natexlab{a}})},\ \Eprint {http://arxiv.org/abs/1412.6226}
  {arXiv:1412.6226 [hep-th]} \BibitemShut {NoStop}%
\bibitem [{\citenamefont {Miyaji}\ and\ \citenamefont
  {Takayanagi}(2015)}]{Miyaji:2015yva}%
  \BibitemOpen
  \bibfield  {author} {\bibinfo {author} {\bibfnamefont {Masamichi}\
  \bibnamefont {Miyaji}}\ and\ \bibinfo {author} {\bibfnamefont {Tadashi}\
  \bibnamefont {Takayanagi}},\ }\bibfield  {title} {\enquote {\bibinfo {title}
  {{Surface/State Correspondence as a Generalized Holography}},}\ }\href
  {\doibase 10.1093/ptep/ptv089} {\bibfield  {journal} {\bibinfo  {journal}
  {PTEP}\ }\textbf {\bibinfo {volume} {2015}},\ \bibinfo {pages} {073B03}
  (\bibinfo {year} {2015})},\ \Eprint {http://arxiv.org/abs/1503.03542}
  {arXiv:1503.03542 [hep-th]} \BibitemShut {NoStop}%
\bibitem [{\citenamefont {Miyaji}\ \emph
  {et~al.}(2015{\natexlab{b}})\citenamefont {Miyaji}, \citenamefont {Numasawa},
  \citenamefont {Shiba}, \citenamefont {Takayanagi},\ and\ \citenamefont
  {Watanabe}}]{Miyaji:2015fia}%
  \BibitemOpen
  \bibfield  {author} {\bibinfo {author} {\bibfnamefont {Masamichi}\
  \bibnamefont {Miyaji}}, \bibinfo {author} {\bibfnamefont {Tokiro}\
  \bibnamefont {Numasawa}}, \bibinfo {author} {\bibfnamefont {Noburo}\
  \bibnamefont {Shiba}}, \bibinfo {author} {\bibfnamefont {Tadashi}\
  \bibnamefont {Takayanagi}}, \ and\ \bibinfo {author} {\bibfnamefont {Kento}\
  \bibnamefont {Watanabe}},\ }\bibfield  {title} {\enquote {\bibinfo {title}
  {{cMERA as Surface/State Correspondence in AdS/CFT}},}\ }\href@noop {} {\
  (\bibinfo {year} {2015}{\natexlab{b}})},\ \Eprint
  {http://arxiv.org/abs/1506.01353} {arXiv:1506.01353 [hep-th]} \BibitemShut
  {NoStop}%
\bibitem [{\citenamefont {'t~Hooft}(1993)}]{'tHooft:1993gx}%
  \BibitemOpen
  \bibfield  {author} {\bibinfo {author} {\bibfnamefont {Gerard}\ \bibnamefont
  {'t~Hooft}},\ }\bibfield  {title} {\enquote {\bibinfo {title} {{Dimensional
  reduction in quantum gravity}},}\ }in\ \href@noop {} {\emph {\bibinfo
  {booktitle} {{Salamfest 1993:0284-296}}}}\ (\bibinfo {year} {1993})\ pp.\
  \bibinfo {pages} {0284--296},\ \Eprint {http://arxiv.org/abs/gr-qc/9310026}
  {arXiv:gr-qc/9310026 [gr-qc]} \BibitemShut {NoStop}%
\bibitem [{\citenamefont {Susskind}(1995)}]{Susskind:1994vu}%
  \BibitemOpen
  \bibfield  {author} {\bibinfo {author} {\bibfnamefont {Leonard}\ \bibnamefont
  {Susskind}},\ }\bibfield  {title} {\enquote {\bibinfo {title} {{The World as
  a hologram}},}\ }\href {\doibase 10.1063/1.531249} {\bibfield  {journal}
  {\bibinfo  {journal} {J. Math. Phys.}\ }\textbf {\bibinfo {volume} {36}},\
  \bibinfo {pages} {6377--6396} (\bibinfo {year} {1995})},\ \Eprint
  {http://arxiv.org/abs/hep-th/9409089} {arXiv:hep-th/9409089 [hep-th]}
  \BibitemShut {NoStop}%
\bibitem [{\citenamefont {Maldacena}(1999)}]{Maldacena:1997re}%
  \BibitemOpen
  \bibfield  {author} {\bibinfo {author} {\bibfnamefont {Juan~Martin}\
  \bibnamefont {Maldacena}},\ }\bibfield  {title} {\enquote {\bibinfo {title}
  {{The Large N limit of superconformal field theories and supergravity}},}\
  }\href {\doibase 10.1023/A:1026654312961} {\bibfield  {journal} {\bibinfo
  {journal} {Int.J.Theor.Phys.}\ }\textbf {\bibinfo {volume} {38}},\ \bibinfo
  {pages} {1113--1133} (\bibinfo {year} {1999})},\ \Eprint
  {http://arxiv.org/abs/hep-th/9711200} {arXiv:hep-th/9711200 [hep-th]}
  \BibitemShut {NoStop}%
\bibitem [{\citenamefont {Vidal}(2007)}]{Vidal:2007hda}%
  \BibitemOpen
  \bibfield  {author} {\bibinfo {author} {\bibfnamefont {G.}~\bibnamefont
  {Vidal}},\ }\bibfield  {title} {\enquote {\bibinfo {title} {{Entanglement
  Renormalization}},}\ }\href {\doibase 10.1103/PhysRevLett.99.220405}
  {\bibfield  {journal} {\bibinfo  {journal} {Phys. Rev. Lett.}\ }\textbf
  {\bibinfo {volume} {99}},\ \bibinfo {pages} {220405} (\bibinfo {year}
  {2007})},\ \Eprint {http://arxiv.org/abs/cond-mat/0512165}
  {arXiv:cond-mat/0512165 [cond-mat]} \BibitemShut {NoStop}%
\bibitem [{\citenamefont {Swingle}(2012)}]{Swingle:2009bg}%
  \BibitemOpen
  \bibfield  {author} {\bibinfo {author} {\bibfnamefont {Brian}\ \bibnamefont
  {Swingle}},\ }\bibfield  {title} {\enquote {\bibinfo {title} {{Entanglement
  Renormalization and Holography}},}\ }\href {\doibase
  10.1103/PhysRevD.86.065007} {\bibfield  {journal} {\bibinfo  {journal} {Phys.
  Rev.}\ }\textbf {\bibinfo {volume} {D86}},\ \bibinfo {pages} {065007}
  (\bibinfo {year} {2012})},\ \Eprint {http://arxiv.org/abs/0905.1317}
  {arXiv:0905.1317 [cond-mat.str-el]} \BibitemShut {NoStop}%
\bibitem [{\citenamefont {Ryu}\ and\ \citenamefont
  {Takayanagi}(2006{\natexlab{a}})}]{Ryu:2006bv}%
  \BibitemOpen
  \bibfield  {author} {\bibinfo {author} {\bibfnamefont {Shinsei}\ \bibnamefont
  {Ryu}}\ and\ \bibinfo {author} {\bibfnamefont {Tadashi}\ \bibnamefont
  {Takayanagi}},\ }\bibfield  {title} {\enquote {\bibinfo {title} {{Holographic
  derivation of entanglement entropy from AdS/CFT}},}\ }\href {\doibase
  10.1103/PhysRevLett.96.181602} {\bibfield  {journal} {\bibinfo  {journal}
  {Phys. Rev. Lett.}\ }\textbf {\bibinfo {volume} {96}},\ \bibinfo {pages}
  {181602} (\bibinfo {year} {2006}{\natexlab{a}})},\ \Eprint
  {http://arxiv.org/abs/hep-th/0603001} {arXiv:hep-th/0603001 [hep-th]}
  \BibitemShut {NoStop}%
\bibitem [{\citenamefont {Ryu}\ and\ \citenamefont
  {Takayanagi}(2006{\natexlab{b}})}]{Ryu:2006ef}%
  \BibitemOpen
  \bibfield  {author} {\bibinfo {author} {\bibfnamefont {Shinsei}\ \bibnamefont
  {Ryu}}\ and\ \bibinfo {author} {\bibfnamefont {Tadashi}\ \bibnamefont
  {Takayanagi}},\ }\bibfield  {title} {\enquote {\bibinfo {title} {{Aspects of
  Holographic Entanglement Entropy}},}\ }\href {\doibase
  10.1088/1126-6708/2006/08/045} {\bibfield  {journal} {\bibinfo  {journal}
  {JHEP}\ }\textbf {\bibinfo {volume} {08}},\ \bibinfo {pages} {045} (\bibinfo
  {year} {2006}{\natexlab{b}})},\ \Eprint {http://arxiv.org/abs/hep-th/0605073}
  {arXiv:hep-th/0605073 [hep-th]} \BibitemShut {NoStop}%
\bibitem [{\citenamefont {Hamilton}\ \emph {et~al.}(2007)\citenamefont
  {Hamilton}, \citenamefont {Kabat}, \citenamefont {Lifschytz},\ and\
  \citenamefont {Lowe}}]{Hamilton:2006fh}%
  \BibitemOpen
  \bibfield  {author} {\bibinfo {author} {\bibfnamefont {Alex}\ \bibnamefont
  {Hamilton}}, \bibinfo {author} {\bibfnamefont {Daniel~N.}\ \bibnamefont
  {Kabat}}, \bibinfo {author} {\bibfnamefont {Gilad}\ \bibnamefont
  {Lifschytz}}, \ and\ \bibinfo {author} {\bibfnamefont {David~A.}\
  \bibnamefont {Lowe}},\ }\bibfield  {title} {\enquote {\bibinfo {title}
  {{Local bulk operators in AdS/CFT: A Holographic description of the black
  hole interior}},}\ }\href {\doibase 10.1103/PhysRevD.75.106001,
  10.1103/PhysRevD.75.129902} {\bibfield  {journal} {\bibinfo  {journal}
  {Phys.Rev.}\ }\textbf {\bibinfo {volume} {D75}},\ \bibinfo {pages} {106001}
  (\bibinfo {year} {2007})},\ \Eprint {http://arxiv.org/abs/hep-th/0612053}
  {arXiv:hep-th/0612053 [hep-th]} \BibitemShut {NoStop}%
\bibitem [{\citenamefont {Kabat}\ \emph {et~al.}(2011)\citenamefont {Kabat},
  \citenamefont {Lifschytz},\ and\ \citenamefont {Lowe}}]{Kabat:2011rz}%
  \BibitemOpen
  \bibfield  {author} {\bibinfo {author} {\bibfnamefont {Daniel}\ \bibnamefont
  {Kabat}}, \bibinfo {author} {\bibfnamefont {Gilad}\ \bibnamefont
  {Lifschytz}}, \ and\ \bibinfo {author} {\bibfnamefont {David~A.}\
  \bibnamefont {Lowe}},\ }\bibfield  {title} {\enquote {\bibinfo {title}
  {{Constructing local bulk observables in interacting AdS/CFT}},}\ }\href
  {\doibase 10.1103/PhysRevD.83.106009} {\bibfield  {journal} {\bibinfo
  {journal} {Phys.Rev.}\ }\textbf {\bibinfo {volume} {D83}},\ \bibinfo {pages}
  {106009} (\bibinfo {year} {2011})},\ \Eprint {http://arxiv.org/abs/1102.2910}
  {arXiv:1102.2910 [hep-th]} \BibitemShut {NoStop}%
\bibitem [{\citenamefont {Kabat}\ and\ \citenamefont
  {Lifschytz}(2014)}]{Kabat:2014kfa}%
  \BibitemOpen
  \bibfield  {author} {\bibinfo {author} {\bibfnamefont {Daniel}\ \bibnamefont
  {Kabat}}\ and\ \bibinfo {author} {\bibfnamefont {Gilad}\ \bibnamefont
  {Lifschytz}},\ }\bibfield  {title} {\enquote {\bibinfo {title} {{Finite N and
  the failure of bulk locality: Black holes in AdS/CFT}},}\ }\href {\doibase
  10.1007/JHEP09(2014)077} {\bibfield  {journal} {\bibinfo  {journal} {JHEP}\
  }\textbf {\bibinfo {volume} {09}},\ \bibinfo {pages} {077} (\bibinfo {year}
  {2014})},\ \Eprint {http://arxiv.org/abs/1405.6394} {arXiv:1405.6394
  [hep-th]} \BibitemShut {NoStop}%
\bibitem [{\citenamefont {Kabat}\ and\ \citenamefont
  {Lifschytz}(2015)}]{Kabat:2015swa}%
  \BibitemOpen
  \bibfield  {author} {\bibinfo {author} {\bibfnamefont {Daniel}\ \bibnamefont
  {Kabat}}\ and\ \bibinfo {author} {\bibfnamefont {Gilad}\ \bibnamefont
  {Lifschytz}},\ }\bibfield  {title} {\enquote {\bibinfo {title} {{Bulk
  equations of motion from CFT correlators}},}\ }\href@noop {} {\  (\bibinfo
  {year} {2015})},\ \Eprint {http://arxiv.org/abs/1505.03755} {arXiv:1505.03755
  [hep-th]} \BibitemShut {NoStop}%
\bibitem [{\citenamefont {Blau}\ \emph {et~al.}(2002)\citenamefont {Blau},
  \citenamefont {Figueroa-O'Farrill}, \citenamefont {Hull},\ and\ \citenamefont
  {Papadopoulos}}]{Blau:2001ne}%
  \BibitemOpen
  \bibfield  {author} {\bibinfo {author} {\bibfnamefont {Matthias}\
  \bibnamefont {Blau}}, \bibinfo {author} {\bibfnamefont {Jose~M.}\
  \bibnamefont {Figueroa-O'Farrill}}, \bibinfo {author} {\bibfnamefont
  {Christopher}\ \bibnamefont {Hull}}, \ and\ \bibinfo {author} {\bibfnamefont
  {George}\ \bibnamefont {Papadopoulos}},\ }\bibfield  {title} {\enquote
  {\bibinfo {title} {{A New maximally supersymmetric background of IIB
  superstring theory}},}\ }\href {\doibase 10.1088/1126-6708/2002/01/047}
  {\bibfield  {journal} {\bibinfo  {journal} {JHEP}\ }\textbf {\bibinfo
  {volume} {01}},\ \bibinfo {pages} {047} (\bibinfo {year} {2002})},\ \Eprint
  {http://arxiv.org/abs/hep-th/0110242} {arXiv:hep-th/0110242 [hep-th]}
  \BibitemShut {NoStop}%
\bibitem [{\citenamefont {Berenstein}\ \emph {et~al.}(2002)\citenamefont
  {Berenstein}, \citenamefont {Maldacena},\ and\ \citenamefont
  {Nastase}}]{Berenstein:2002jq}%
  \BibitemOpen
  \bibfield  {author} {\bibinfo {author} {\bibfnamefont {David~Eliecer}\
  \bibnamefont {Berenstein}}, \bibinfo {author} {\bibfnamefont {Juan~Martin}\
  \bibnamefont {Maldacena}}, \ and\ \bibinfo {author} {\bibfnamefont
  {Horatiu~Stefan}\ \bibnamefont {Nastase}},\ }\bibfield  {title} {\enquote
  {\bibinfo {title} {{Strings in flat space and pp waves from N=4
  superYang-Mills}},}\ }\href {\doibase 10.1088/1126-6708/2002/04/013}
  {\bibfield  {journal} {\bibinfo  {journal} {JHEP}\ }\textbf {\bibinfo
  {volume} {0204}},\ \bibinfo {pages} {013} (\bibinfo {year} {2002})},\ \Eprint
  {http://arxiv.org/abs/hep-th/0202021} {arXiv:hep-th/0202021 [hep-th]}
  \BibitemShut {NoStop}%
\bibitem [{\citenamefont {Das}\ \emph {et~al.}(2002)\citenamefont {Das},
  \citenamefont {Gomez},\ and\ \citenamefont {Rey}}]{Das:2002cw}%
  \BibitemOpen
  \bibfield  {author} {\bibinfo {author} {\bibfnamefont {Sumit~R.}\
  \bibnamefont {Das}}, \bibinfo {author} {\bibfnamefont {Cesar}\ \bibnamefont
  {Gomez}}, \ and\ \bibinfo {author} {\bibfnamefont {Soo-Jong}\ \bibnamefont
  {Rey}},\ }\bibfield  {title} {\enquote {\bibinfo {title} {{Penrose limit,
  spontaneous symmetry breaking and holography in PP wave background}},}\
  }\href {\doibase 10.1103/PhysRevD.66.046002} {\bibfield  {journal} {\bibinfo
  {journal} {Phys.Rev.}\ }\textbf {\bibinfo {volume} {D66}},\ \bibinfo {pages}
  {046002} (\bibinfo {year} {2002})},\ \Eprint
  {http://arxiv.org/abs/hep-th/0203164} {arXiv:hep-th/0203164 [hep-th]}
  \BibitemShut {NoStop}%
\bibitem [{\citenamefont {Kiritsis}\ and\ \citenamefont
  {Pioline}(2002)}]{Kiritsis:2002kz}%
  \BibitemOpen
  \bibfield  {author} {\bibinfo {author} {\bibfnamefont {E.}~\bibnamefont
  {Kiritsis}}\ and\ \bibinfo {author} {\bibfnamefont {B.}~\bibnamefont
  {Pioline}},\ }\bibfield  {title} {\enquote {\bibinfo {title} {{Strings in
  homogeneous gravitational waves and null holography}},}\ }\href {\doibase
  10.1088/1126-6708/2002/08/048} {\bibfield  {journal} {\bibinfo  {journal}
  {JHEP}\ }\textbf {\bibinfo {volume} {0208}},\ \bibinfo {pages} {048}
  (\bibinfo {year} {2002})},\ \Eprint {http://arxiv.org/abs/hep-th/0204004}
  {arXiv:hep-th/0204004 [hep-th]} \BibitemShut {NoStop}%
\bibitem [{\citenamefont {Leigh}\ \emph {et~al.}(2002)\citenamefont {Leigh},
  \citenamefont {Okuyama},\ and\ \citenamefont {Rozali}}]{Leigh:2002pt}%
  \BibitemOpen
  \bibfield  {author} {\bibinfo {author} {\bibfnamefont {Robert~G.}\
  \bibnamefont {Leigh}}, \bibinfo {author} {\bibfnamefont {Kazumi}\
  \bibnamefont {Okuyama}}, \ and\ \bibinfo {author} {\bibfnamefont {Moshe}\
  \bibnamefont {Rozali}},\ }\bibfield  {title} {\enquote {\bibinfo {title} {{P
  P waves and holography}},}\ }\href {\doibase 10.1103/PhysRevD.66.046004}
  {\bibfield  {journal} {\bibinfo  {journal} {Phys.Rev.}\ }\textbf {\bibinfo
  {volume} {D66}},\ \bibinfo {pages} {046004} (\bibinfo {year} {2002})},\
  \Eprint {http://arxiv.org/abs/hep-th/0204026} {arXiv:hep-th/0204026 [hep-th]}
  \BibitemShut {NoStop}%
\bibitem [{\citenamefont {Berenstein}\ and\ \citenamefont
  {Nastase}(2002)}]{Berenstein:2002sa}%
  \BibitemOpen
  \bibfield  {author} {\bibinfo {author} {\bibfnamefont {David}\ \bibnamefont
  {Berenstein}}\ and\ \bibinfo {author} {\bibfnamefont {Horatiu}\ \bibnamefont
  {Nastase}},\ }\bibfield  {title} {\enquote {\bibinfo {title} {{On light cone
  string field theory from superYang-Mills and holography}},}\ }\href@noop {}
  {\  (\bibinfo {year} {2002})},\ \Eprint {http://arxiv.org/abs/hep-th/0205048}
  {arXiv:hep-th/0205048 [hep-th]} \BibitemShut {NoStop}%
\bibitem [{\citenamefont {Bianchi}\ \emph {et~al.}(1998)\citenamefont
  {Bianchi}, \citenamefont {Green}, \citenamefont {Kovacs},\ and\ \citenamefont
  {Rossi}}]{Bianchi:1998nk}%
  \BibitemOpen
  \bibfield  {author} {\bibinfo {author} {\bibfnamefont {Massimo}\ \bibnamefont
  {Bianchi}}, \bibinfo {author} {\bibfnamefont {Michael~B.}\ \bibnamefont
  {Green}}, \bibinfo {author} {\bibfnamefont {Stefano}\ \bibnamefont {Kovacs}},
  \ and\ \bibinfo {author} {\bibfnamefont {Giancarlo}\ \bibnamefont {Rossi}},\
  }\bibfield  {title} {\enquote {\bibinfo {title} {{Instantons in
  supersymmetric Yang-Mills and D instantons in IIB superstring theory}},}\
  }\href {\doibase 10.1088/1126-6708/1998/08/013} {\bibfield  {journal}
  {\bibinfo  {journal} {JHEP}\ }\textbf {\bibinfo {volume} {08}},\ \bibinfo
  {pages} {013} (\bibinfo {year} {1998})},\ \Eprint
  {http://arxiv.org/abs/hep-th/9807033} {arXiv:hep-th/9807033 [hep-th]}
  \BibitemShut {NoStop}%
\bibitem [{\citenamefont {Verlinde}(2015)}]{Verlinde:2015qfa}%
  \BibitemOpen
  \bibfield  {author} {\bibinfo {author} {\bibfnamefont {Herman}\ \bibnamefont
  {Verlinde}},\ }\bibfield  {title} {\enquote {\bibinfo {title} {{Poking Holes
  in AdS/CFT: Bulk Fields from Boundary States}},}\ }\href@noop {} {\
  (\bibinfo {year} {2015})},\ \Eprint {http://arxiv.org/abs/1505.05069}
  {arXiv:1505.05069 [hep-th]} \BibitemShut {NoStop}%
\bibitem [{\citenamefont {Nakayama}\ and\ \citenamefont
  {Ooguri}(2015)}]{Nakayama:2015mva}%
  \BibitemOpen
  \bibfield  {author} {\bibinfo {author} {\bibfnamefont {Yu}~\bibnamefont
  {Nakayama}}\ and\ \bibinfo {author} {\bibfnamefont {Hirosi}\ \bibnamefont
  {Ooguri}},\ }\bibfield  {title} {\enquote {\bibinfo {title} {{Bulk Locality
  and Boundary Creating Operators}},}\ }\href@noop {} {\  (\bibinfo {year}
  {2015})},\ \Eprint {http://arxiv.org/abs/1507.04130} {arXiv:1507.04130
  [hep-th]} \BibitemShut {NoStop}%
\bibitem [{\citenamefont {Metsaev}\ and\ \citenamefont
  {Tseytlin}(2002)}]{Metsaev:2002re}%
  \BibitemOpen
  \bibfield  {author} {\bibinfo {author} {\bibfnamefont {R.~R.}\ \bibnamefont
  {Metsaev}}\ and\ \bibinfo {author} {\bibfnamefont {Arkady~A.}\ \bibnamefont
  {Tseytlin}},\ }\bibfield  {title} {\enquote {\bibinfo {title} {{Exactly
  solvable model of superstring in Ramond-Ramond plane wave background}},}\
  }\href {\doibase 10.1103/PhysRevD.65.126004} {\bibfield  {journal} {\bibinfo
  {journal} {Phys. Rev.}\ }\textbf {\bibinfo {volume} {D65}},\ \bibinfo {pages}
  {126004} (\bibinfo {year} {2002})},\ \Eprint
  {http://arxiv.org/abs/hep-th/0202109} {arXiv:hep-th/0202109 [hep-th]}
  \BibitemShut {NoStop}%
\bibitem [{\citenamefont {Sadri}\ and\ \citenamefont
  {Sheikh-Jabbari}(2004)}]{Sadri:2003pr}%
  \BibitemOpen
  \bibfield  {author} {\bibinfo {author} {\bibfnamefont {Darius}\ \bibnamefont
  {Sadri}}\ and\ \bibinfo {author} {\bibfnamefont {Mohammad~M.}\ \bibnamefont
  {Sheikh-Jabbari}},\ }\bibfield  {title} {\enquote {\bibinfo {title} {{The
  Plane wave / superYang-Mills duality}},}\ }\href {\doibase
  10.1103/RevModPhys.76.853} {\bibfield  {journal} {\bibinfo  {journal}
  {Rev.Mod.Phys.}\ }\textbf {\bibinfo {volume} {76}},\ \bibinfo {pages} {853}
  (\bibinfo {year} {2004})},\ \Eprint {http://arxiv.org/abs/hep-th/0310119}
  {arXiv:hep-th/0310119 [hep-th]} \BibitemShut {NoStop}%
\bibitem [{\citenamefont {Gaberdiel}\ and\ \citenamefont
  {Green}(2003)}]{Gaberdiel:2002hh}%
  \BibitemOpen
  \bibfield  {author} {\bibinfo {author} {\bibfnamefont {Matthias~R.}\
  \bibnamefont {Gaberdiel}}\ and\ \bibinfo {author} {\bibfnamefont
  {Michael~B.}\ \bibnamefont {Green}},\ }\bibfield  {title} {\enquote {\bibinfo
  {title} {{The D instanton and other supersymmetric D branes in 2B plane wave
  string theory}},}\ }\href {\doibase 10.1016/S0003-4916(03)00082-4} {\bibfield
   {journal} {\bibinfo  {journal} {Annals Phys.}\ }\textbf {\bibinfo {volume}
  {307}},\ \bibinfo {pages} {147--194} (\bibinfo {year} {2003})},\ \Eprint
  {http://arxiv.org/abs/hep-th/0211122} {arXiv:hep-th/0211122 [hep-th]}
  \BibitemShut {NoStop}%
\bibitem [{\citenamefont {Bergman}\ \emph {et~al.}(2003)\citenamefont
  {Bergman}, \citenamefont {Gaberdiel},\ and\ \citenamefont
  {Green}}]{Bergman:2002hv}%
  \BibitemOpen
  \bibfield  {author} {\bibinfo {author} {\bibfnamefont {Oren}\ \bibnamefont
  {Bergman}}, \bibinfo {author} {\bibfnamefont {Matthias~R.}\ \bibnamefont
  {Gaberdiel}}, \ and\ \bibinfo {author} {\bibfnamefont {Michael~B.}\
  \bibnamefont {Green}},\ }\bibfield  {title} {\enquote {\bibinfo {title}
  {{D-brane interactions in type IIB plane wave background}},}\ }\href
  {\doibase 10.1088/1126-6708/2003/03/002} {\bibfield  {journal} {\bibinfo
  {journal} {JHEP}\ }\textbf {\bibinfo {volume} {0303}},\ \bibinfo {pages}
  {002} (\bibinfo {year} {2003})},\ \Eprint
  {http://arxiv.org/abs/hep-th/0205183} {arXiv:hep-th/0205183 [hep-th]}
  \BibitemShut {NoStop}%
\bibitem [{\citenamefont {Skenderis}\ and\ \citenamefont
  {Taylor}(2002)}]{Skenderis:2002vf}%
  \BibitemOpen
  \bibfield  {author} {\bibinfo {author} {\bibfnamefont {Kostas}\ \bibnamefont
  {Skenderis}}\ and\ \bibinfo {author} {\bibfnamefont {Marika}\ \bibnamefont
  {Taylor}},\ }\bibfield  {title} {\enquote {\bibinfo {title} {{Branes in AdS
  and p p wave space-times}},}\ }\href {\doibase 10.1088/1126-6708/2002/06/025}
  {\bibfield  {journal} {\bibinfo  {journal} {JHEP}\ }\textbf {\bibinfo
  {volume} {0206}},\ \bibinfo {pages} {025} (\bibinfo {year} {2002})},\ \Eprint
  {http://arxiv.org/abs/hep-th/0204054} {arXiv:hep-th/0204054 [hep-th]}
  \BibitemShut {NoStop}%
\bibitem [{\citenamefont {Miyaji}\ \emph
  {et~al.}(2015{\natexlab{c}})\citenamefont {Miyaji}, \citenamefont {Numasawa},
  \citenamefont {Shiba}, \citenamefont {Takayanagi},\ and\ \citenamefont
  {Watanabe}}]{MIyaji:2015mia}%
  \BibitemOpen
  \bibfield  {author} {\bibinfo {author} {\bibfnamefont {Masamichi}\
  \bibnamefont {Miyaji}}, \bibinfo {author} {\bibfnamefont {Tokiro}\
  \bibnamefont {Numasawa}}, \bibinfo {author} {\bibfnamefont {Noburo}\
  \bibnamefont {Shiba}}, \bibinfo {author} {\bibfnamefont {Tadashi}\
  \bibnamefont {Takayanagi}}, \ and\ \bibinfo {author} {\bibfnamefont {Kento}\
  \bibnamefont {Watanabe}},\ }\bibfield  {title} {\enquote {\bibinfo {title}
  {{Gravity Dual of Quantum Information Metric}},}\ }\href@noop {} {\
  (\bibinfo {year} {2015}{\natexlab{c}})},\ \Eprint
  {http://arxiv.org/abs/1507.07555} {arXiv:1507.07555 [hep-th]} \BibitemShut
  {NoStop}%
\bibitem [{\citenamefont {Lashkari}\ and\ \citenamefont
  {Van~Raamsdonk}(2015)}]{Lashkari:2015hha}%
  \BibitemOpen
  \bibfield  {author} {\bibinfo {author} {\bibfnamefont {Nima}\ \bibnamefont
  {Lashkari}}\ and\ \bibinfo {author} {\bibfnamefont {Mark}\ \bibnamefont
  {Van~Raamsdonk}},\ }\bibfield  {title} {\enquote {\bibinfo {title}
  {{Canonical Energy is Quantum Fisher Information}},}\ }\href@noop {} {\
  (\bibinfo {year} {2015})},\ \Eprint {http://arxiv.org/abs/1508.00897}
  {arXiv:1508.00897 [hep-th]} \BibitemShut {NoStop}%
\bibitem [{\citenamefont {Kim}\ and\ \citenamefont {Yee}(2003)}]{Kim:2002tj}%
  \BibitemOpen
  \bibfield  {author} {\bibinfo {author} {\bibfnamefont {Nakwoo}\ \bibnamefont
  {Kim}}\ and\ \bibinfo {author} {\bibfnamefont {Jung-Tay}\ \bibnamefont
  {Yee}},\ }\bibfield  {title} {\enquote {\bibinfo {title} {{Supersymmetry and
  branes in M theory plane waves}},}\ }\href {\doibase
  10.1103/PhysRevD.67.046004} {\bibfield  {journal} {\bibinfo  {journal} {Phys.
  Rev.}\ }\textbf {\bibinfo {volume} {D67}},\ \bibinfo {pages} {046004}
  (\bibinfo {year} {2003})},\ \Eprint {http://arxiv.org/abs/hep-th/0211029}
  {arXiv:hep-th/0211029 [hep-th]} \BibitemShut {NoStop}%
\end{thebibliography}%
\end{document}